\documentclass[conference]{IEEEtran}
\ifCLASSINFOpdf
\else
\fi
\usepackage{amsmath}
\usepackage{amssymb}
\usepackage{booktabs}
\usepackage{url}
\usepackage{subcaption}

\usepackage{graphicx}
\usepackage{bm}

\usepackage{balance}

\usepackage[most]{tcolorbox}
\usepackage{xcolor}
\usepackage{tabularx}
\usepackage{fontawesome} 
\usepackage{diagbox}

\usepackage{enumitem}

\newtcolorbox{userbox}{
  colback=gray!10,
  colframe=gray!30,
  boxrule=0.3pt,
  arc=2pt,
  left=4pt,
  right=4pt,
  top=4pt,
  bottom=4pt,
  before skip=4pt,
  after skip=4pt,
  width=\linewidth,
  boxsep=4pt
}

\newtcolorbox{botbox}{
  colback=blue!5,
  colframe=blue!30,
  boxrule=0.3pt,
  arc=2pt,
  left=4pt,
  right=4pt,
  top=4pt,
  bottom=4pt,
  before skip=4pt,
  after skip=4pt,
  width=\linewidth,
  boxsep=4pt
}

\usepackage{marvosym}

\begin{document}
%
\title{ThinkTrap: Denial-of-Service Attacks against Black-box LLM Services via Infinite Thinking}

	

%

\IEEEoverridecommandlockouts

\author{\IEEEauthorblockN{Yunzhe Li\IEEEauthorrefmark{1},
Jianan Wang\IEEEauthorrefmark{1},
Hongzi Zhu\IEEEauthorrefmark{1}\textsuperscript{\Letter}\thanks{\textsuperscript{\Letter} Hongzi Zhu is the corresponding author of this paper.}, 
James Lin\IEEEauthorrefmark{1}, Shan Chang\IEEEauthorrefmark{2}  and
Minyi Guo\IEEEauthorrefmark{1}}
\IEEEauthorblockA{\IEEEauthorrefmark{1}Shanghai Jiao Tong University, \IEEEauthorrefmark{2}Donghua University}
\IEEEauthorblockA{\{yunzhe.li, divinenoah, hongzi, james\}@sjtu.edu.cn, changshan@dhu.edu.cn, guo-my@cs.sjtu.edu.cn}}


\IEEEoverridecommandlockouts
\makeatletter\def\@IEEEpubidpullup{6.5\baselineskip}\makeatother
\IEEEpubid{\parbox{\columnwidth}{
		Network and Distributed System Security (NDSS) Symposium 2026\\
		23-27 February 2026, San Diego, CA, USA\\
		ISBN 979-8-9894372-8-3\\
		https://dx.doi.org/10.14722/ndss.2026.24639\\
		www.ndss-symposium.org
}
\hspace{\columnsep}\makebox[\columnwidth]{}}

\maketitle

\begin{abstract}
Large Language Models (LLMs) have become foundational components in a wide range of applications, including natural language understanding and generation, embodied intelligence, and scientific discovery. As their computational requirements continue to grow, these models are increasingly deployed as cloud-based services, allowing users to access powerful LLMs via the Internet. However, this deployment model introduces a new class of threat: denial-of-service (DoS) attacks via unbounded reasoning, where adversaries craft specially designed inputs that cause the model to enter excessively long or infinite generation loops. These attacks can exhaust backend compute resources, degrading or denying service to legitimate users. To mitigate such risks, many LLM providers adopt a closed-source, black-box setting to obscure model internals. In this paper, we propose ThinkTrap, a novel input-space optimization framework for DoS attacks against LLM services even in black-box environments. The core idea of ThinkTrap is to first map discrete tokens into a continuous embedding space, then undertake efficient black-box optimization in a low-dimensional subspace exploiting input sparsity. The goal of this optimization is to identify adversarial prompts that induce extended or non-terminating generation across several state-of-the-art LLMs, achieving DoS with minimal token overhead. We evaluate the proposed attack across multiple commercial, closed-source LLM services. Our results demonstrate that, even far under the restrictive request frequency limits commonly enforced by these platforms, typically capped at ten requests per minute (10 RPM), the attack can degrade service throughput to as low as 1\% of its original capacity, and in some cases, induce complete service failure.

\end{abstract}


%
\IEEEpeerreviewmaketitle

\section{Introduction}

Large Language Models (LLMs) have emerged as a transformative foundation for modern AI systems, enabling powerful capabilities such as natural language understanding and generation \cite{sanderson2023gpt} \cite{qiu2024llm}, embodied intelligence \cite{mon2025embodied} \cite{li2024embodied}, and scientific discovery \cite{zheng2025large} \cite{truhn2023large}. 
Due to their massive computational demands, especially during multi-step inference or long-form generation, LLMs are increasingly deployed as cloud-based services to serve a broad and growing user base. However, this introduces a critical vulnerability, \emph{i.e.}, denial-of-service (DoS) attacks \cite{schuba1997analysis} \cite{pelechrinis2010denial} that exploit the recursive reasoning process of an LLM. Unlike conventional DoS attacks that flood the network or overwhelm server endpoints, these newer attacks introduce intensive computation costs by inducing LLMs to \emph{think} endlessly or generate prohibitively long outputs \cite{dong25}. 
One single malicious input can monopolize substantial GPU time, queue slots, or memory resources, effectively starving legitimate users and causing service degradation or outright outages \cite{shumailov2021sponge}. This asymmetric threat, where a small token input leads to unbounded computation at cloud servers, represents a novel and particularly insidious attack surface in the era of large-scale AI deployment.

Launching a DoS attack against closed-source LLMs must meet the following requirements. First, the attack should only rely on the input-output interface of an LLM service, which exposes quite limited information with no internal information such as logits or attention weights. Second, the attack must be cost-efficient because attackers also need to pay for LLM queries. As a result, effective adversarial prompts must be generated with minimal API calls. Third, the attack must be robust across models and potential defenses. Modern LLMs often include safeguards such as output filters or truncation mechanisms. Successful attacks must generalize across these variations and remain effective despite unknown internal changes.

Previous attack attempts to induce long or non-terminating outputs from LLMs can be broadly classified into three categories, \emph{i.e.}, semantic-based \cite{kumar2025overthinking} \cite{chen2024not}, gradient-based \cite{dong25} \cite{rico16}, and heuristic-based \cite{feng2024llmeffichecker} \cite{shumailov2021sponge}. The first category employs semantic manipulation, such as presenting the model with inherently open-ended prompts or complex tasks (\emph{e.g.}, Olympiad-level mathematics problems) to encourage extended responses \cite{kumar2025overthinking}. While occasionally successful, these techniques often lack robustness and generalizability, typically relying on fragile prompt engineering and exhibiting effectiveness only on specific models. The second category utilizes gradient-based optimization methods, commonly aiming to suppress the probability of generating end-of-sequence (EoS) token in order to prolong output \cite{dong25}. Although effective in white-box settings, such approaches necessitate access to internal model parameters or output logits, rendering them unsuitable for use with proprietary or closed-source LLM APIs. Finally, the third category involves heuristic-driven search strategies at the token level to identify input prompts that lead to extended outputs \cite{shumailov2021sponge}. These methods, however, tend to be computationally inefficient and incur high query costs, limiting their scalability and feasibility in practice. As a result, existing approaches are constrained by limitations in generality, efficiency, and applicability to black-box scenarios.

In this paper, we propose an attack framework, called ThinkTrap, which can conduct a DoS attack on closed-source black-box LLM service. The core idea of ThinkTrap is to employ the derivative-free optimization of input tokens with respect to the output length, even under the constraint that the LLM autoregressive generation process is non-differentiable and only provides limited black-box access to input-output pairs. By approximating the direction in which an input prompt elongates the model's output, ThinkTrap efficiently searches for adversarial prompts that maximize generation length, thereby amplifying the computational burden on the LLM service and inducing a denial-of-service (DoS) effect. To this end, the ThinkTrap design encounters two main challenges as follows. 

First, the input space of large language models (LLMs) is inherently discrete, consisting of sequences of tokens, which poses a significant obstacle to the application of derivative-free optimization methods that typically operate over continuous domains. In contrast to continuous spaces, where infinitesimal perturbations produce smooth changes in objective functions, minor modifications to token sequences can induce abrupt and non-monotonic variations in model behavior. This discreteness severely limits the stability and efficiency of direct optimization in the original input space. To address this challenge, we introduce a continuous surrogate space in which optimization can be more effectively conducted. Specifically, we map discrete token sequences, \emph{i.e.}, prompts, into a continuous embedding space, which serves as a proxy domain for exploration. Within this space, we apply derivative-free optimization strategies to identify directions that are likely to induce longer or more computationally intensive outputs from the target LLM. The optimized embedding vectors are then projected back to the closest valid token sequences using a nearest-neighbor decoding mechanism, ensuring compatibility with the actual input requirements of victim LLM service.
This surrogate-based formulation enables derivative-free prompt optimization while respecting the discrete structure of natural language inputs. It provides a stable and tractable framework for inducing adversarial behavior in black-box language models deployed as cloud services.

Second, derivative-free optimization methods are inherently less efficient than gradient-based approaches, especially in high-dimensional input spaces. In the case of LLMs, the dimensionality of the optimization space can be extremely large. For example, when optimizing over a prompt of 100 tokens with a typical LLM, \emph{i.e.}, LLaMA-70B, the search space spans over 400k dimensions. Exploring such a vast space using derivative-free methods can be prohibitively expensive and slow to converge.
To tackle this issue, we exploit a key property of large language models that their response behavior often lies in a low intrinsic dimensionality subspace \cite{armen21} \cite{qin2024exploring}. Prior studies and our empirical observations suggest that modifying only a small subset of the parameters of LLMs is often sufficient to induce significant changes in model behavior \cite{xu24lora} \cite{sun2022black}. Leveraging this insight, we design a low-dimensional optimization strategy that constrains the subgradient search to a small number of editable token positions. By selecting and optimizing only a few strategically chosen input tokens while keeping the rest fixed, we effectively reduce the search space by orders of magnitude. This dimensionality reduction drastically improves the efficiency of our attack without sacrificing effectiveness.

To evaluate the effectiveness and generality of ThinkTrap, we conduct extensive real-world experiments across a wide range of popular closed-source LLM APIs, including services based on GPT, DeepSeek, and Gemini families. Our results demonstrate that ThinkTrap achieves a high attack success rate in black-box settings, consistently identifying prompts that induce excessively long outputs and impose significant computational burdens on the target services.
Moreover, to further investigate the impact of the attack, we deploy a high-performance LLM service on a private server equipped with 16 Ascend GPUs and emulate ThinkTrap-style attacks under controlled conditions. Experimental results indicate that even a low-rate adversarial query stream, \emph{e.g.}, issuing only five requests per minute, can significantly degrade service quality when prompts are crafted using our method. Specifically, the attack saturates GPU memory and computational resources, leading to up to a 100$\times$ increase in response latency, a reduction in throughput to as low as 1\% of the original performance, and, in extreme cases, complete service failure due to GPU exhaustion. These findings confirm that ThinkTrap can serve as an effective denial-of-service (DoS) attack method against both commercial LLM services and self-hosted LLM deployments.
The main contributions made in this paper are highlighted as follows:

\begin{itemize}
    \item We identify a new DoS vulnerability in the closed-source black-box LLM services through prompt-level manipulation, even without access to internal model gradients or parameters.
    \item We propose ThinkTrap, a novel attack framework that leverages subgradient-guided optimization in a continuous surrogate space to craft adversarial prompts that elicit abnormally long responses, significantly increasing the target model’s computational load.
    \item We validate ThinkTrap through extensive experiments on both public LLM services and private LLM deployments. The results demonstrate that ThinkTrap can induce substantial degradation in service performance, highlighting a new class of realistic threats to large-scale LLM systems.
\end{itemize}

\section{Related Work}

\subsection{DoS attack on Machine Learning Services}

Several recent studies have demonstrated the feasibility of DoS attacks on machine learning systems through the use of computationally intensive inputs. Sponge Example \cite{shumailov2021sponge} is among the first to show that adversarial inputs can significantly increase energy consumption and inference latency in neural networks, effectively degrading model availability. Later work reveals that uniform inputs and sparse activations can exacerbate this effect, suggesting that ``sponge behavior'' is not limited to worst-case optimization but may also emerge under structured perturbations \cite{muller2024impact}. These findings highlight a class of resource exhaustion attacks that exploit the computational characteristics of deep models rather than their prediction accuracy.

More recent efforts have adapted this threat model to object detection and autonomous systems. Phantom Sponges \cite{shapira2023phantom} exploit inefficiencies in the non-maximum suppression (NMS) to inflate the number of processed detections, and follow-up work \cite{schoof2024beyond} has enhanced these attacks by introducing multi-modal perturbations that intensify computational demand. Meanwhile, SlowTrack \cite{ma2024slowtrack} and SlowLiDAR \cite{liu2023slowlidar} have shown that imperceptible input modifications can substantially increase latency in camera- and LiDAR-based perception systems, respectively. These techniques reveal a broader attack surface where adversarial examples can compromise real-time guarantees in safety-critical applications by targeting system responsiveness rather than correctness.


\subsection{DoS Attacks on LLMs}

LLMs are susceptible to DoS attacks due to their substantial model size and the autoregressive nature of their decoding process, which results in inference costs scaling linearly with the length of the generated output. Consequently, adversaries can launch DoS attacks by inducing the model to produce excessively long or complex outputs \cite{shumailov2021sponge} \cite{chen2022nmtsloth} \cite{feng2024llmeffichecker}. Existing DoS attack methods, however, predominantly target open-source LLMs. For encoder-decoder architectures, attacks exploit cross-attention mechanisms by compressing numerous tokens into a single input sequence, thereby increasing computational burden \cite{shumailov2021sponge}. Such strategies are ineffective against decoder-only models, which lack cross-attention modules. Perturbation-based approaches aim to modify tokens critical to output length \cite{feng2024llmeffichecker}, but the widespread adoption of Byte-Pair Encoding (BPE) tokenization \cite{rico16}, which enhances LLMs’ tolerance to typographical errors, has largely diminished their effectiveness. Gradient-based optimization techniques \cite{dong25} attempt to craft adversarial prompts by minimizing the probability of generating the end-of-sequence (EoS) token. However, these methods require access to token-level probabilities, which are typically unavailable in popular LLM APIs. Semantic-based attacks leverage complex input prompts, such as Olympiad-level mathematics problems, to provoke longer outputs \cite{kumar2025overthinking} \cite{ geiping2024coercing}, but these approaches suffer from instability due to their dependence on specific model behaviors.

\subsection{Security Threats to Black-box LLMs}

Security concerns surrounding black-box access to LLMs have grown substantially. Black-box tuning has demonstrated that commercial LLMs can be adapted to downstream tasks without access to model gradients, by leveraging zeroth-order optimization and query-efficient strategies \cite{sun2022black}. Similarly, PromptBoosting shows that accurate black-box text classification can be achieved with as few as 10 forward passes, revealing the adaptability of LLMs to prompt-based inference despite strict API request constraints \cite{hou2023promptboosting}. These approaches, though designed for benign applications, inadvertently highlight the susceptibility of black-box LLMs to repeated probing and exploitation.

More adversarial efforts have shown that universal and transferable prompt-based attacks can reliably bypass safety alignment mechanisms across tasks, even in black-box scenarios \cite{zou2023universal}. A large-scale evaluation of jailbreak attacks versus defenses further indicates that aligned safety mechanisms remain fragile when facing adaptive prompt manipulation, particularly in real-world LLM deployments \cite{XuLDLP24}. Additionally, Cold-Attack introduces a novel jailbreak strategy that combines stealthiness and controllability, making detection and mitigation significantly more difficult for LLM service providers \cite{GuoYZQ024}. These efforts collectively reveal that LLM services expose a broad and under-defended attack surface. Building on these insights, we present a new threat: a denial-of-service (DoS) attack vector uniquely enabled by the access of black-box LLMs, which exploits the model’s resource consumption behavior to impair availability without requiring any internal knowledge or cooperation.


\section{System and Attack Model}

\subsection{Background on LLM Inference}

Large language models (LLMs) based on the Transformer architecture  generate text in an autoregressive manner. Inference typically consists of two stages \cite{mindie2023}:

\begin{itemize}
    \item \textbf{Prefill stage}: The model processes the entire input prompt in one forward pass to initialize key-value (KV) cache representations. This stage has computational cost proportional to the input length.
    \item \textbf{Decode stage}: Tokens are then generated one-by-one. For each output token, the model performs a full forward pass over the cached context, making the cost of generation grow approximately linearly with the output length.
\end{itemize}

Because each token requires querying large parameter matrices and maintaining KV cache on memory-intensive accelerators (\emph{e.g.}, GPUs), inference cost, latency, and memory footprint scale with both prompt and output lengths. As a result, unusually long responses can substantially increase resource usage, slow down concurrent requests, and incur higher operational cost in real deployments. This computational structure motivates our study of attacks that intentionally trigger extremely long model outputs.

\subsection{Attacker}

The attacker has black-box access to the LLM service via its API. Their capabilities include:

\begin{itemize}
    \item \textbf{Query Access}: The attacker can issue arbitrary text prompts to the model and monitor the corresponding outputs, including the length of the generated responses.
    \item \textbf{No Internal Access}: Owing to the high deployment cost or the proprietary nature of the LLMs, the attacker is unable to access the model’s parameters, gradients, architecture, or confidence scores.
    \item \textbf{Budget Constraints}: The attacker has limited budgets for the number of tokens consumed during the search for the attack prompts, as token usage incurs cost when interacting with the LLM API. Thus, attackers aim to maximize output length per token spent.
\end{itemize}

\begin{figure}
    \centering
    \includegraphics[width=\linewidth]{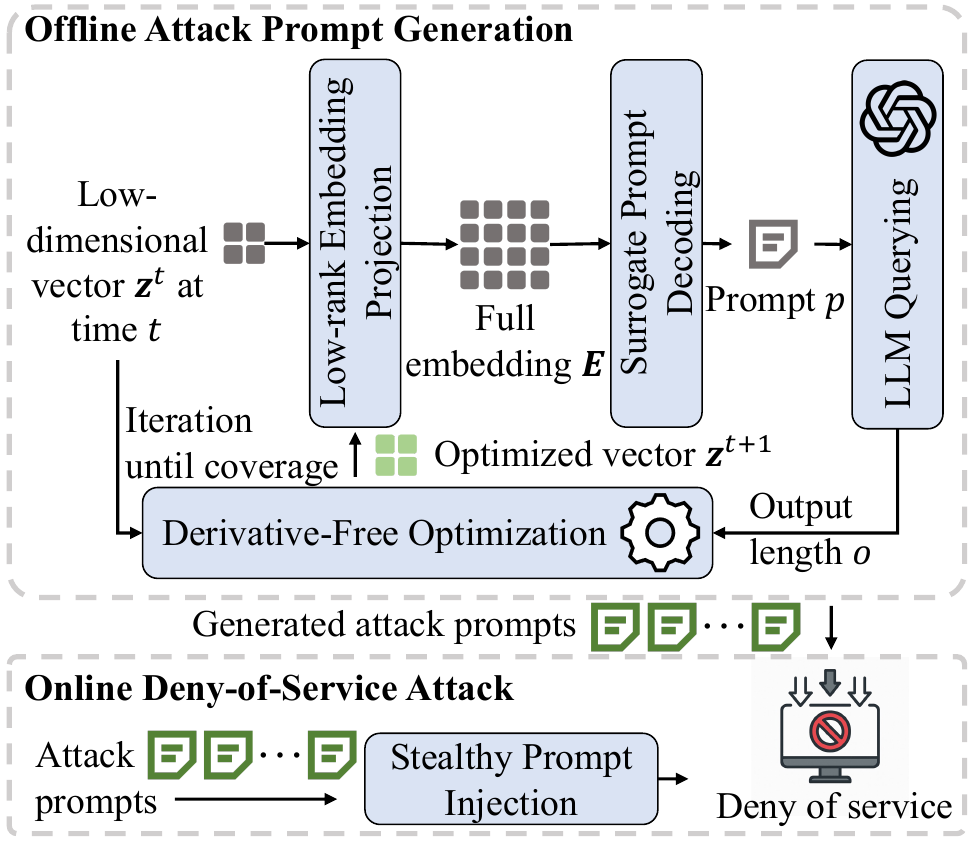}
    \caption{Attack overview of the proposed ThinkTrap system, where the attack prompts are first generated offline and then injected into the LLM in a stealthy way to conduct a denial-of-service attack.}
    \label{fig:overview}
\end{figure}

\subsection{Victim}

The victim is an LLM service providing inference through a black-box API with the following characteristics:

\begin{itemize}
    \item \textbf{Input-Output Interface}: The victim provides a publicly exposed interface that accepts discrete textual inputs from arbitrary users over the internet and returns the corresponding generated text responses.
    \item \textbf{Resource Constraints}: Although LLMs are typically deployed in cloud environments, their large model sizes and autoregressive decoding \cite{vaswani2017attention} make inference inherently expensive. In particular, generating each output token requires a full forward pass, causing computational cost, memory consumption, and latency to grow roughly linearly with the output length. As output grows longer, the demand on shared GPU resources increases significantly, which can lead to higher operational costs and service delays, especially under sustained or concurrent requests.
    \item \textbf{Basic Defenses}: The victim may implement basic safeguards such as input length restrictions, rate limiting, and token quotas to constrain resource usage. These measures are generally effective against naive attacks that rely on sending a large number of requests concurrently. For example, commercial LLM APIs like GPT-4 enforce strict request limits (\emph{e.g.}, no more than 10 requests per minute). However, such defenses still permit low-rate, sustained queries for normal service.
\end{itemize}

\section{Overview of ThinkTrap}


The proposed ThinkTrap system consists of two stages as illustrated in Figure~\ref{fig:overview}, namely offline attack prompt generation and online denial-of-service attack, with all notations summarized in Table~\ref{tab:notation}. At a high level, ThinkTrap operates entirely through a query–response loop in a strict black-box setting where the attacker can only submit textual prompts and observe the generated outputs. To search for effective prompts under these constraints, ThinkTrap optimizes a low-dimensional latent vector that is projected into an embedding, decoded into a discrete prompt using a surrogate vocabulary, and evaluated solely through the output length returned by the victim API. This scalar feedback guides a derivative-free optimization process that gradually improves the latent vector. The optimized prompts obtained offline are then issued online at a moderate rate to trigger excessively long generations and degrade service availability.

\begin{table}[]
\centering
\caption{Summary of notations.}
\begin{tabular}{ll}
\toprule
Notation & Description \\
\midrule
$\bm{p}^{\text{att}}$ & Attack prompts to generate \\
$t$ & Index of the current optimization iteration \\
$p^t$ & Prompt generated at the $t$-th optimization iteration \\
$\bm{z}^t$ & Low-dimensional vector of the prompt $p^t$ \\
$m$ & Dimensionality of each token vector in $\bm{z}^t$ \\
$\bm{E}^t$ & Full embedding of the prompt $p^t$ \\
$d$ & Dimensionality of each token embedding in $\bm{E}^t$ \\
$\bm{A}$ & Random projection matrix \\
$o^t$ & Generated output length in tokens \\
$\bm{T}^{\text{sur.}}$ & Embedding table of the surrogate encoder \\
$w_j$ & $j$-th word token in surrogate encoder $\bm{T}^{\text{sur.}}$ \\
$\bm{T}_j^{\text{sur.}}$ & Embedding of word token $w_j$ in $\bm{T}^{\text{sur.}}$ \\
$\mathcal{M}^\text{vic}$ & Black-box victim LLM \\
\bottomrule
\end{tabular}
\label{tab:notation}
\end{table}

\textbf{Offline Attack Prompt Generation (APG).} Given access to the target LLM via its API, the attacker first generates attack prompts $\bm{p}^{\text{att}}$ offline using a derivative-free optimization approach as detailed below.


\emph{1) Low-rank Embedding Projection (LEP).} Given the low-dimensional vector, denoted as $\bm{z}^t$ at time $t$, the LEP module projects it into the full embedding, denoted as $\bm{E}^t$ via a random projection matrix $\bm{A}$, \emph{i.e.}, $\bm{E}^t = \bm{A}\bm{z}^t$.

\emph{2) Surrogate Prompt
Decoding (SPD).} The SPD module maps the full embedding $\bm{E}^t$ back to a textual prompt $p^t$ by decoding it via nearest-neighbor token retrieval on a surrogate prompt encoding space to approximate the target LLM’s encoding process in the black-box scenario.

\emph{3) LLM Querying (LQ).} The LQ module evaluates the current prompt $p^t$ by querying the LLM through its API and assessing the length of the generated output as an indicator of the attack efficacy $o^t$ of the prompt $p^t$.

\emph{4) Derivative-Free Optimization (DFO).} To enhance the effectiveness of the attack prompt, the low-dimensional latent vector $\bm{z}^t$ is iteratively optimized using a derivative-free method, guided by the observed efficacy score $o^t$ of the corresponding prompt $p^t$. The updated vector, denoted as $\bm{z}^{t+1}$, serves as the basis for the next optimization step. This process repeats until convergence or until a successful attack is achieved.

\textbf{Online Deny-of-Service Attack (DSA).} The attacker leverages the offline-generated attack prompts $\bm{p}^{\text{att}}$ and injects them into the LLM service through the public LLM API in a stealthy manner, aiming to evade defense detection while conducting a denial-of-service (DoS) attack.

\section{Offline Attack Prompt Generation}

\subsection{Low-rank Embedding Projection}

Optimizing the prompt embedding $\bm{E}^t$ in the full continuous space $\mathbb{R}^{L \times d}$ is computationally prohibitive due to its high dimensionality. For example, with a prompt length $L = 100$ and embedding dimension $d = 4096$ (\emph{e.g.}, LLaMA-2-70B \cite{touvron2023llama}), the optimization space contains over 400K parameters, rendering derivative-free methods inefficient because they scale poorly with dimensionality.
To mitigate this, we introduce a low-dimensional latent vector $\bm{z}^t \in \mathbb{R}^m$ ($m \ll Ld$) and use a fixed, randomly initialized projection matrix $\bm{A} \in \mathbb{R}^{(Ld)\times m}$ to map the latent representation to the full embedding space, \emph{i.e.}, $\bm{E}^t = \bm{A}\bm{z}^t$.
The optimization is then performed over the low-dimensional vector $\bm{z}^t$ to improve optimization efficiency.
This design exploits the redundancy and sparsity of LLM embedding spaces \cite{armen21, qin2024exploring}, enabling efficient search in a compact subspace. We then first introduce the design of the projection matrix $\bm{A}$ and introduce the embedding projection.

\subsubsection{Projection Matrix Construction}


The design of $\bm{A}$ must meet several important criteria to ensure that the search in the low-dimensional latent space remains effective and unbiased:
\begin{itemize}
    \item The projected directions should be isotropic, meaning they do not favor any particular axis in the high-dimensional space;
    \item The mapping should avoid amplifying specific coordinates, ensuring that the sampling process is balanced;
    \item The projection should maintain diversity, such that different latent vectors $\bm{z}^t$ produce sufficiently distinct embeddings $\bm{E}^t$ to allow exploration of a broad set of adversarial candidates.
\end{itemize}
To satisfy these properties, we construct $\bm{A}$ with entries sampled independently from a Gaussian distribution \cite{zhang2016gaussian}:
\begin{equation}
    A_{i,j} \sim \mathcal{N}(0, \frac{1}{m}),
\end{equation}
where $A_{i,j}$ denotes the value of $i$-th line and $j$-column in the matrix $\bm{A}$, and $m$ denotes the target embedding dimension of the low-dimensional vector $\bm{z}$. 

\subsubsection{Embedding Projection}

Given the latent vector $\bm{z}^t \in \mathbb{R}^m$ and a fixed projection matrix $\bm{A} \in \mathbb{R}^{Ld \times m}$, the prompt embedding is directly obtained via a linear transformation: 
\begin{equation}
    \bm{E}^t = \bm{A} \bm{z}^t \in \mathbb{R}^{Ld}.
\end{equation}
This formulation allows the optimization to take place in a compact latent space, substantially reducing the parameter search space. The use of a random Gaussian projection for $\bm{A}$ promotes subspace diversity and helps retain the expressive capacity of the original embedding space.

\subsection{Surrogate Prompt
Decoding}

To decode the embedding $\bm{E}^t$ to the prompt $p^t$, we apply a core decoding strategy to map the optimized continuous embeddings back into discrete token sequences via nearest neighbor search in the model’s token embedding space. This step is essential in black-box settings, where public LLM APIs only accept textual inputs. Ideally, the decoding should be performed using the target model’s token embedding matrix to ensure semantic alignment. However, in black-box scenarios, this internal decoder is inaccessible. To this end, we leverage an important empirical property of large language models: \emph{the token embedding spaces across different models exhibit a high degree of alignment}, owing to shared tokenization schemes, overlapping training corpora, and convergent training dynamics \cite{brown2020language, gurnee2023language}. This enables us to use a surrogate model’s decoder as a practical substitute for nearest neighbor decoding. In practice, we observe that such surrogate-based decoding achieves strong performance, despite the lack of access to the target model’s embedding layer.

Formally, given a prompt embedding $\bm{E}^t$, to convert this into a discrete prompt $p^t = (w^t_1, \dots, w^t_L)$, where $w^t_i$ denotes the $i$-th word token in the prompt, we perform nearest neighbor decoding using a publicly available surrogate token embedding matrix $\bm{T}^{\text{sur.}} \in \mathbb{R}^{|\mathcal{V}| \times d}$, where each embedding $\bm{T}_j^{\text{sur.}}$ represents the embedding of word token $w_j$. For each token position $i$, we compute:
\begin{equation}
w_i = \arg\min_{j \in \mathcal{V}} \|\bm{e}_i^t - \bm{T}^{\text{sur.}}_j\|_2,
\end{equation}
where $\bm{e}_i$ denotes the $i$-th embedding in $\bm{E}^t$. The resulting sequence $p^t = (w^t_1, \dots, w^t_L)$ is then submitted to the black-box LLM API for evaluation.

\subsection{LLM Querying}

Given a discrete prompt $p$, we submit it to the target language model via its public API. Although the optimization process operates in the latent space $\bm{z}$ and performs intermediate computations in the continuous embedding space $\bm{E}$, neither representation is directly compatible with the API interface, which only accepts tokenized textual inputs. Therefore, only the decoded prompt $p$ can be used for querying the model.

Formally, we invoke the LLM with $p^t$ as input and evaluate its output to obtain the attack objective $o^t$, such as the number of generated tokens:
\begin{equation}
o^t = \mathcal{M}^{\text{vic}}(p^t),
\end{equation}
where $\mathcal{M}^{\text{vic}}$ denotes the black-box victim model being attacked. This black-box querying process forms the only observable channel through which the attacker can evaluate and optimize the attack objective.

\subsection{Derivative-Free Optimization}

To optimize the latent vector $\bm{z}$ in a black-box setting, we adopt a derivative-free optimization (DFO) strategy. Unlike gradient-based methods, which rely on access to model parameters or backpropagation, DFO methods require only the evaluation of the objective function, making them particularly well-suited for black-box attacks on large language models (LLMs) via public APIs. In our context, the objective function assesses the effectiveness of a given latent vector $\bm{z}$ by converting it into a discrete prompt and measuring generation-based attack metrics, \emph{i.e.}, the length of the model’s output.

Formally, the optimization problem is defined as:
\begin{equation}
\max_{\bm{z}^t \in \mathbb{R}^m} \mathcal{L}(\bm{z}^t) = o^t,
\end{equation}
where $\mathcal{L}(\bm{z})$ denotes the scalar objective value (\emph{i.e.}, generation length) returned by the victim model $\mathcal{M}^{\text{vic}}$ in response to the prompt derived from $\bm{z}$.

To solve this problem, we employ the Covariance Matrix Adaptation Evolution Strategy (CMA-ES) \cite{auger2012tutorial}, a state-of-the-art DFO algorithm that maintains a multivariate Gaussian search distribution $\mathcal{N}(\bm{\mu}^{(t)}, \bm{\Sigma}^{(t)})$ over the latent space. The distribution parameters are iteratively updated based on the performance of sampled candidates with the following steps.

\subsubsection{Initialization}

At iteration $t = 0$, the search is initialized with mean vector $\bm{\mu}^{0} = \bm{0}$ and isotropic covariance matrix $\bm{\Sigma}^{0} = \sigma^2 \bm{I}$, where $\sigma > 0$ controls the initial search radius. This defines an initial uniform search distribution over the latent space.

\subsubsection{Optimization Procedure}

At each iteration $t$, CMA-ES proceeds as follows:

\textbf{Sampling.} A population of $N$ latent candidates ${ \bm{z}_i^{t} }$, $i=1, 2, \cdots, N$, is drawn from the current distribution:
\begin{equation}
\bm{z}_i^{t} \sim \mathcal{N}(\bm{\mu}^{t}, \bm{\Sigma}^{t}).
\end{equation}

\textbf{Evaluation.} Each candidate $\bm{z}_i^{t}$ is projected to an embedding $\bm{E}_i^t = \bm{A} \bm{z}_i^{t}$, decoded into a discrete prompt $p_i$, and submitted to the victim LLM $\mathcal{M}^{\text{vic}}$ to obtain its corresponding objective score $o^t_1, o^t_2, \cdots, o^t_N$.

\textbf{Selection and Recombination.} The candidates are ranked according to their scores ${ o_i^t }$, and the top $k$ individuals are selected. A new mean vector is computed via a weighted average of these top-performing candidates:
\begin{equation}
\bm{\mu}^{(t+1)} = \sum_{j=1}^k w_j \bm{z}_{j}^{t},
\end{equation}
where $\bm{z}_{j}^{t}$ denotes the $j$-th best candidate and ${ w_j }$ are predefined positive weights summing to 1.

\textbf{Covariance Update.} The covariance matrix is updated to reflect the empirical distribution of the selected candidates:
\begin{equation}
\bm{\Sigma}^{t+1} = \sum_{j=1}^{k} w_j (\bm{z}_{j}^{t} - \bm{\mu}^{t+1})(\bm{z}_{j}^{t} - \bm{\mu}^{t+1})^T + \epsilon \bm{I},
\end{equation}
where $\epsilon$ is a small constant added for numerical stability.

This update mechanism captures both the principal directions and the variance of successful candidates, enabling the algorithm to adaptively explore high-performing regions of the latent space. The optimization proceeds until convergence or a predefined querying budget is reached.

By directly operating on the latent vector $\bm{z}$ and relying solely on black-box evaluations of $\mathcal{M}^{\text{vic}}$, this framework enables efficient search for adversarial prompts without requiring access to model gradients or internal parameters.

\section{Online Denial-of-Service Attack}

Modern large language models (LLMs), especially those deployed via public APIs, are typically equipped with basic security mechanisms designed to defend against abuse and misuse. These include input filtering, rate limiting, behavioral detection, and generation constraints \cite{wu2024new}, all of which pose significant challenges for sustained or high-frequency adversarial querying. Consequently, launching an effective deny-of-service (DoS) attack in such settings requires carefully evading these protections while still delivering adversarial prompts capable of degrading model availability or utility.

To this end, we design an online attack framework that incrementally submits crafted prompts at a moderate query rate, in order to bypass rate limiting and avoid triggering abuse detection mechanisms. Our approach does not rely on overwhelming the system with high-throughput requests. It exploits the fact that carefully optimized prompts can consume excessive model computation, even when submitted infrequently, by inducing long and resource-intensive generations. This results in a form of \emph{slow} DoS, which imposes sustained computational burden on the model over time.

Concretely, we assume the target LLM service enforces a minimum interval of $t$ seconds between accepted queries from a single user. To comply with this constraint and remain undetected, the attacker submits attack prompts at a fixed rate no faster than once every $t$ seconds. Each prompt is selected from a pool of previously optimized attack prompts.

\section{Evaluation}

\subsection{Methodology}

\subsubsection{Implementation}

We develop a prototype implementation of the attack based on a Python CMA-ES optimization library~\footnote{\url{https://github.com/CyberAgent/cmaes}}. The attack is conducted under a black-box setting by interacting with target LLMs through the unified Model Router API~\footnote{\url{https://openrouter.ai/}}, which dispatches requests to various backend models and returns their responses. Due to computational constraints, we set the maximum generation length to 4096 tokens, a commonly adopted upper bound in LLM decoding \cite{dong25}, and sufficient to observe generation collapse or abnormal length behaviors. The input prompt length is fixed at 20 tokens to balance optimization efficiency and input compactness, while the decoding temperature is set to its default value of 1.0 to reflect the standard sampling configuration used in public-facing LLM APIs.

\subsubsection{LLMs}


\begin{table}[t]
\centering
\caption{
Overview of evaluated LLMs. ``OSS'' denotes whether the model is open-source.
``Tokens/wk'' denotes the weekly token consumption during June 23–29, 2025.
``Price'' represents the approximate cost per 1M \textit{output} tokens for public API inference at the time of evaluation. 
Values are intended to contextualize practical deployment scale and cost rather than serve as exact billing references.
}
\label{tab:llm}
\begin{tabular}{lllllll}
\toprule
Model & Params & Provider & OSS & Tokens/wk & Price$^\dagger$ \\ 
\midrule
Gemini 2.5 Pro   & N/A     & Google     & No  & 88.8B  & \$10 \\
Lumimaid         & 70B     & NeverSleep & No  & 12.4M  & \$3 \\
Magistral        & N/A     & Mistral    & No  & 58.7M  & \$5 \\
GPT-o4           & N/A     & OpenAI     & No  & 3.22B  & \$4.4 \\
MAI DS R1        & 671B    & Microsoft  & Yes & 998M   & \$1.2 \\
DS Qwen3         & 8B      & DeepSeek   & Yes & 1.93B  & \$0.02 \\
Llama 3.2        & 3B      & Meta       & Yes & 10.4B  & \$0.02 \\
DS R1            & 671B    & DeepSeek   & Yes & 63.2B  & \$2.15 \\
\bottomrule
\end{tabular}
\begin{flushleft}
\footnotesize{$^\dagger$ Open-source models (MAI DS R1, DS Qwen3, Llama 3.2, DS R1) are accessible via third-party providers offering free usage tiers for platform promotion. As a result, attackers may leverage such free services to conduct attacks without incurring any monetary cost.}
\end{flushleft}
\end{table}

We evaluate a diverse set of eight LLMs, as listed in Table \ref{tab:llm}, covering both proprietary and open-source models. This selection includes frontier-scale models such as DeepSeek R1 (671B) and widely used close-sourced LLMs like GPT o4 and Gemini 2.5 Pro. The weekly token usage reflects their real-world adoption in our experiments conducted from June 23 to 29, 2025.

\subsubsection{Baselines}

We consider the following four black-box baselines for evaluation.

\begin{itemize}
    \item \textbf{Decoy Problem} \cite{kumar2025overthinking}: We collect a set of 20 samples by prompting GPT-4o to generate a set of open-ended, high-complexity questions spanning a wide range of domains, including physics, machine learning, economics, biology, and philosophy. These questions are intentionally selected to induce prolonged reasoning and multi-step generation, making them particularly challenging for LLMs.
    
    \item \textbf{Semantic Problem} \cite{dong25}: We enhance the decoy problems by adding explicit semantic cues that encourage longer responses. Specifically, each question is rephrased to include instructions such as “Output a longer explanation” or “Provide a more detailed discussion,” thereby guiding the language model to extend its generation.
    
    \item \textbf{LLMEffiChecker} \cite{feng2024llmeffichecker}: We adopt the attack proposed in LLMEffiChecker, which performs word-level perturbations. Specifically, it first measures the impact of each word on the output length and identifies the word that contributes most to shortening the generation. Then, it randomly substitutes this word to induce longer outputs from the model.
    
    \item \textbf{Sponge Examples} \cite{shumailov2021sponge}: This approach leverages a genetic algorithm to perform word-level optimization. By iteratively evolving input sequences, it generates adversarial examples that induce excessive computational cost and prolonged output generation in language models.
\end{itemize}

\begin{figure*}[]
    \centering
    
    \begin{subfigure}[]{0.49\linewidth}
        \centering
        \includegraphics[width=\linewidth]{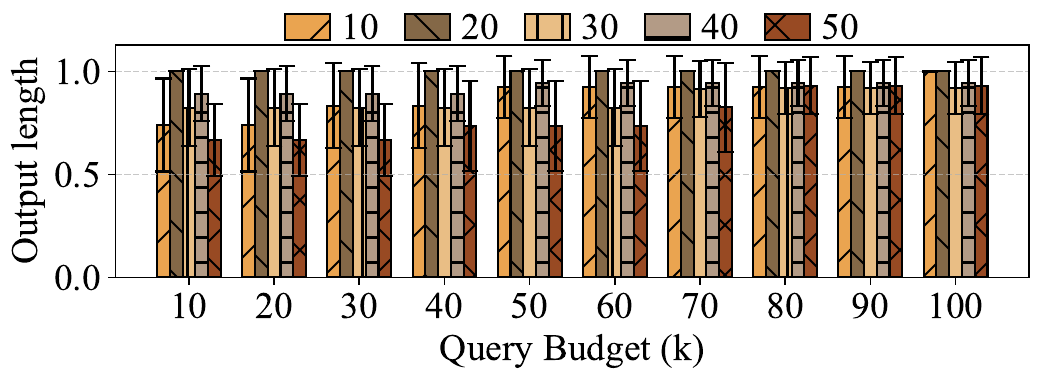}
        \caption{Different prompt lengths.}
        \label{fig:prompt_length}
    \end{subfigure}
    \hfill
    \begin{subfigure}[]{0.49\linewidth}
        \centering
        \includegraphics[width=\linewidth]{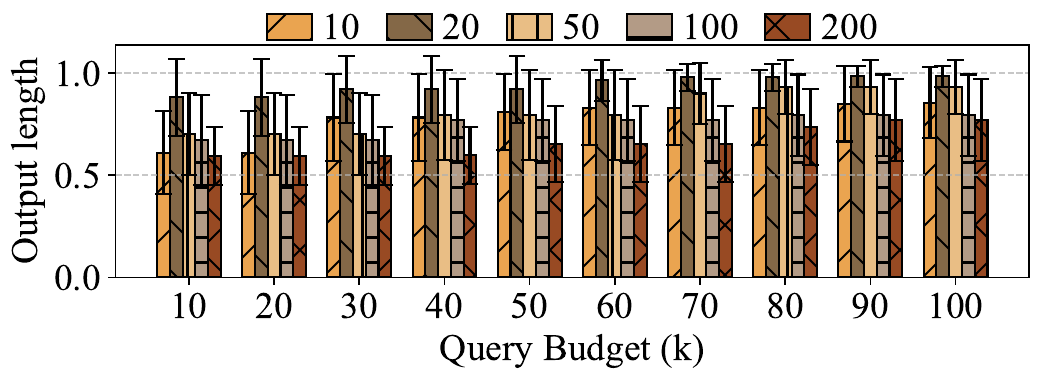}
        \caption{Different latent vector dimensions.}
        \label{fig:prompt_dim}
    \end{subfigure}
    \caption{Output length of DeepSeek R1 with respect to the upper bound of 4096 on ThinkTrap under (a) different prompt lengths and (b) different latent vector dimensions, where a non-monotonic trend can be observed in both hyperparameters for a balance of prompt expressiveness and search efficiency.}
    \label{fig:length_prompt}
\end{figure*}

\subsubsection{Metrics}

We consider the metrics in terms of LLM behaviors in various aspects.

\begin{itemize}
    \item \textbf{LLM output length}: Output length measures the total number of tokens generated in response to a generated attack prompt. This reflects how effectively the prompt can elicit prolonged generation. Considering the possibility that a large language model (LLM) may generate unbounded output when engaged in infinite reasoning, we impose a maximum output length of 4096 tokens during evaluation. For consistency and clarity in presentation, all reported output lengths are normalized with respect to 4096 tokens.
    \item \textbf{Tokens per second (TPS)}: Tokens per second, \emph{i.e.}, TPS, captures the generation throughput, defined as the number of output tokens produced per second. A lower throughput under attack inputs indicates degraded model efficiency and increased inference cost.

    \item \textbf{Time to first token (TTFT)}: Time to first token, \emph{i.e.}, TTFT, refers to the time elapsed between submitting the prompt and receiving the first generated token. Increased latency may suggest that the prompt induces heavier computational burden during decoding initialization.

    \item \textbf{GPU memory consumption}: GPU memory consumption records the peak GPU memory usage during model inference. Prompts that trigger abnormally high memory usage can be indicative of resource exhaustion vulnerabilities.

\end{itemize}

\subsection{Hyperparameter Search}

\subsubsection{Prompt Length}

We first investigate the effect of prompt length on the attack results. To this end, we plot the result of normalized output length with five input prompt lengths on DeepSeek R1, \emph{i.e.,} 10, 20, 30, 40, 50, in Fig. \ref{fig:prompt_length}. 
We observe a non-monotonic trend in the effectiveness of prompt-based attacks under a fixed query budget in Fig. \ref{fig:prompt_length}. As prompt length increases, the attack initially becomes more potent due to enhanced expressiveness. However, beyond a certain length (\emph{e.g.}, 20), performance degrades. This is because longer prompts expand the search space, causing the derivative-free optimization to waste more budget on uninformative directions. Consequently, the probability of discovering highly effective adversarial prompts within a fixed number of queries decreases.
As a result, we choose a balanced prompt length of 20 for the best attack performance on a given budget.

\subsubsection{Latent Vector Dimension}




We then investigate the effect of subspace dimensions on the attack results. Specifically, we report the normalized output length under five latent dimensions, \emph{i.e.}, 10, 20, 50, 100, and 200, each averaged over 20 independent trials, as shown in Fig.~\ref{fig:prompt_dim}. The results exhibit a non-monotonic trend. Moderate dimensionalities enhance attack effectiveness by providing a richer search space and greater expressive capacity for adversarial prompts. However, once the dimension becomes sufficiently large, performance begins to degrade, due to increased optimization difficulty and the curse of dimensionality. Accordingly, we select a latent dimension of 20 as a practical choice that balances achievable output length and optimization efficiency.

\subsection{LLM Output Length of Attack Prompts}

\begin{figure*}[]
    \centering
    \includegraphics[width=\linewidth]{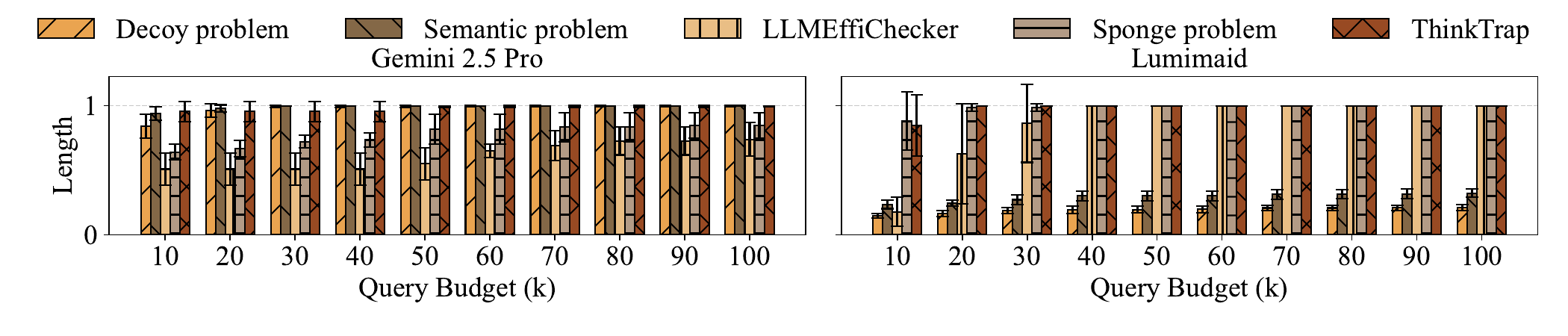}
    \includegraphics[width=\linewidth]{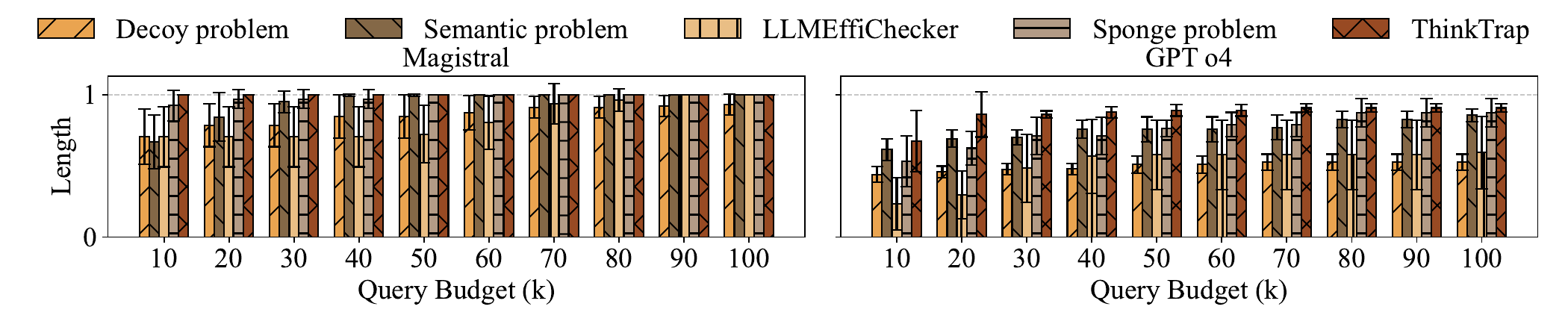}
    \includegraphics[width=\linewidth]{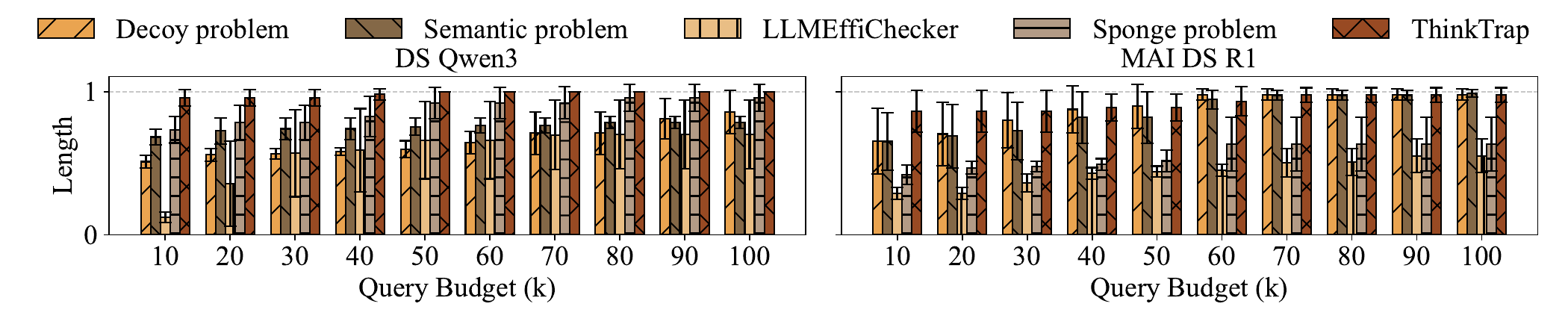}
    \includegraphics[width=\linewidth]{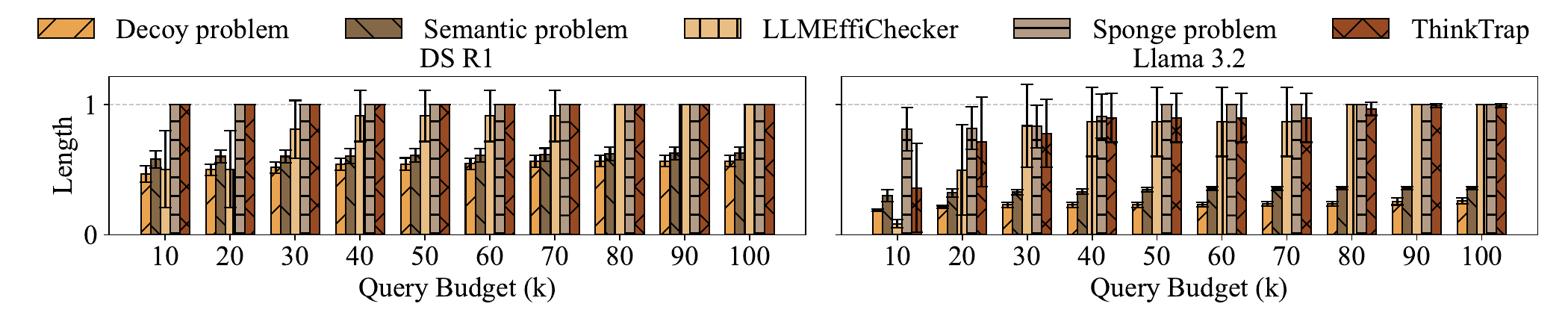}

    \caption{Output length of the evaluated eight LLMs on ThinkTrap and all the four baselines with respect to the upper bound of 4096, where different baseline methods exhibit varying performance across different models, but ThinkTrap consistently achieves the highest output length across all LLMs. The advantage of ThinkTrap is particularly evident under lower generation budgets, demonstrating its efficiency in maximizing output with minimal resources.}
    \label{fig:length}

\end{figure*}

To evaluate the generation efficiency of different LLMs under varying output budgets, we conduct experiments by setting the allowed output budget from 10K to 100K tokens. For each budget level, we compare different attack methods in terms of their ability to generate adversarial prompts that induce the longest possible model outputs. To mitigate the impact of randomness, each experiment is repeated five times. Fig. \ref{fig:length} reports the average output lengths along with standard deviations across the five runs.


\subsubsection{Comparison with Baselines}

We first compare ThinkTrap with existing baselines to show its superior performance.

\textbf{ThinkTrap succeeds even on models where semantic-based attacks fail.} We observe that the performance of attack methods based on semantics, \emph{i.e.}, decoy problem and semantic problem, heavily depends on the specific architecture and alignment strategy of the target LLM. While these methods may achieve moderate success on certain instruction-tuned models, they often fail to generalize across models with different pretraining objectives or decoding behaviors, \emph{e.g.}, Lumimaid, DS R1, Llama 3.2. This suggests that purely semantic manipulations lack the robustness and universality required for cross-model transferability.

\textbf{ThinkTrap attains long outputs with far lower query budgets than heuristic-driven baselines, which struggle under limited budgets.} We can see that heuristic-driven search methods, \emph{i.e.}, LLMEffiChecker and Sponge problem, do not perform well, especially on Gemini 2.5 Pro and MAI DS R1 when budget is low. This is because these approaches typically rely on manually designed scoring functions or rule-based mutation strategies. As a result, they require significantly more queries or larger generation budgets to discover effective adversarial prompts. This inefficiency becomes particularly pronounced when operating under tight computational constraints, limiting their practicality for large-scale attacks. In contrast, ThinkTrap leverages an adaptive optimization strategy, enabling faster convergence with fewer queries.

\textbf{ThinkTrap demonstrates robust performance across seven of the eight evaluated LLMs, with particularly strong advantages under limited query budgets.}
These results indicate its capability to efficiently exploit generation dynamics under resource constraints. Unlike baselines that typically depend on substantial query budgets to induce long outputs, ThinkTrap can discover prompts that trigger disproportionately long responses with limited tokens. While its performance is consistently strong for seven out of eight models, we observe that for LLama 3.2 at very low budgets, Sponge achieves higher output lengths. This exception highlights the importance of model-specific factors, but overall, ThinkTrap's generalization across architectures and budget levels underscores its practical value in realistic denial-of-service settings where adversaries may face strict cost constraints. Its efficiency arises from an adaptive lightweight search strategy that transfers effectively across models with different alignment and decoding characteristics.


\subsubsection{Attack Effects on Various LLMs}

We then analyze the attack effect of ThinkTrap on various types of LLMs.

\textbf{ThinkTrap maintains strong effectiveness across both closed-source and open-source LLMs, indicating that its performance does not depend on access to model internals.} We can see the ThinkTrap perform well on both closed-source and open-source LLMs. This is because ThinkTrap is a black-box attack method without requiring access to model internals or gradient information, which further underscores its practicality for real-world adversarial scenarios.

\textbf{ThinkTrap remains effective across diverse model families, whereas baseline methods show less consistent performance across architectures.} We can see that ThinkTrap performs well on various model families including Gemini, Luminmaid, Magistral, GPT, Qwen, DeepSeek, and Llama. This shows that modern LLMs all suffer from this kind of attacks. Despite that different attack methods attack different parts of LLM, ThinkTrap can outperform other baselines on every evaluated model family, showing the high importance of the proposed ThinkTrap method.

\begin{figure}
    \centering
    \includegraphics[width=\linewidth]{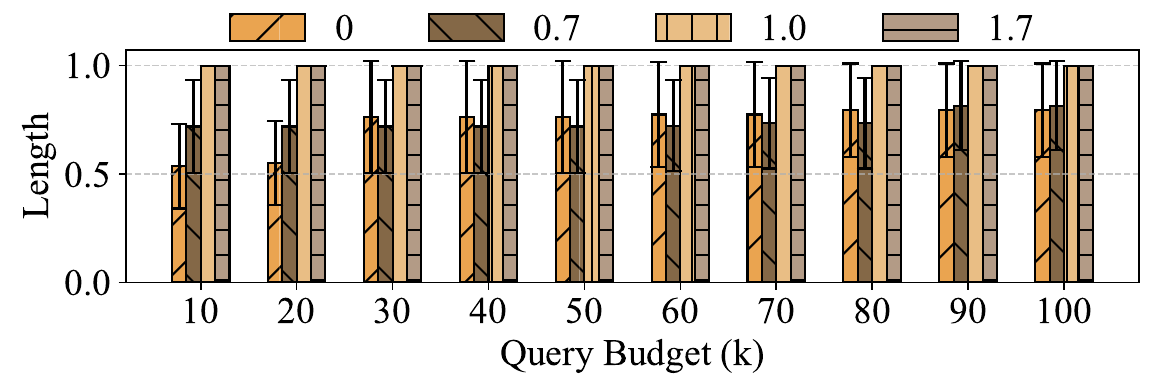}
    \caption{Output length relative to the maximum limit of 4096 tokens for the eight evaluated LLMs under ThinkTrap, across varying decoding temperatures (\emph{i.e.}, 0, 0.7, 1, 1.7), where higher temperatures, introducing greater sampling randomness, consistently result in longer outputs.}
    \label{fig:temp}

\end{figure}




\textbf{ThinkTrap exhibits strong effectiveness across LLMs of varying scales, indicating that its performance generalizes well regardless of model size.} We can also see that LLMs of various mainstream sizes, \emph{i.e.}, 3B (Llama 3.2), 8B (DS Qwen 3), 70B (Lumimaid), and 671B (DS R1 and MAI DS R1). Moreover, their behaviors do not vary despite different sizes. This shows that despite the inference cost various, their risks under the DoS attack equal. In other words, larger LLMs are not safer.

\textbf{ThinkTrap is effective on both thinking and non-thinking LLMs, suggesting that vulnerability to the attack extends beyond models with explicit reasoning features.} We can see that ThinkTrap can also work well on the non-thinking base LLM, \emph{i.e.}, Llama 3.2. This shows that non-thinking LLM may also suffer from the DoS attack. While we fail to attack many early published LLMs such as Llama 2. We owe this failure to the fact that early LLMs even do not have an ability to output a long enough results. However, current LLMs such as Llama-3.2 can generate longer outputs than early models, making them more vulnerable to this attack.

\subsubsection{Attack Effects on Various Decoding Strategies}

Decoding temperature is an important parameter which affects the diversity of decoding behavior. We also plot the performance of ThinkTrap on DS R1 under different decoding temperatures in Fig. \ref{fig:temp} to show its behavior of various temperatures.

\textbf{ThinkTrap attack remains effective across a wide range of decoding temperature settings.}
Decoding temperature is a hyperparameter that controls the stochasticity of token sampling and thereby regulates output diversity in LLM decoding.
Fig.~\ref{fig:temp} illustrates that the attack consistently succeeds across both low and high temperature configurations.
A low decoding temperature yields low output diversity, causing the model to generate repetitive lexical or token patterns that sustain the attack. 
In contrast, higher temperatures increase output variability, making the model less inclined to produce the $\texttt{<EOS>}$ token.
In most LLMs, $\texttt{<EOS>}$ acts as a high-logit, low-entropy token near the end of a sentence, where it signals sequence completion.
Raising the temperature weakens this token’s dominance, resulting in longer and more chaotic continuations, during which the model may enter a self-reinforcing or perpetual thinking loop.

\subsubsection{Attack Cost Analysis}

We further quantify the attack cost of different methods to assess the practical feasibility of the proposed attack.

\textbf{ThinkTrap imposes a very small cost and can be executed adaptively given inferred system capabilities.} In particular, ThinkTrap attains a successful attack (\emph{i.e.}, forcing the model to produce 4k+ tokens) with an attack budget of only 10k tokens on Gemini 2.5 Pro, Magistral, DS Qwen3, and DS R1. In these cases, end-to-end execution requires only a few minutes. For other evaluated LLMs, a comparable success is also achievable within 100k tokens, corresponding to runtimes below one hour. Using the prices reported in Table~\ref{tab:llm}, the monetary cost of such attacks is negligible in practice. A representative cost for DS R1 requiring a 10k-token budget is only \$0.0215. Even for relatively costly services, the expense remains small, \emph{e.g.}, approximately \$0.10 for Gemini 2.5 Pro requiring a 10k-token budget, and \$0.44 for GPT o4 requiring a 100k-token budget. These results demonstrate that ThinkTrap is both low-cost and practically feasible in realistic  scenarios.

\begin{figure*}
    \centering
    \includegraphics[width=\linewidth]{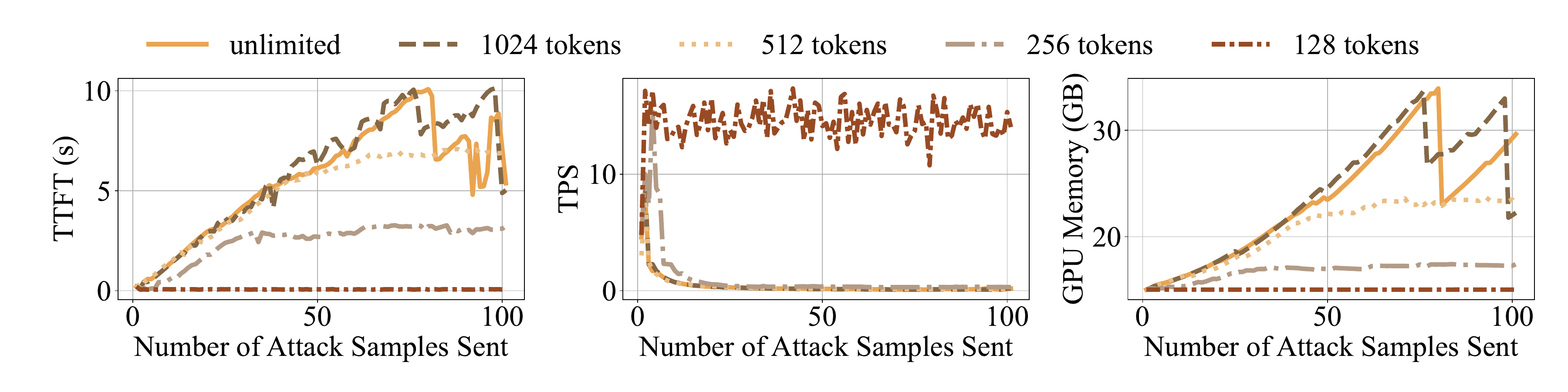}
    \caption{Impact of ThinkTrap attack on the DeepSeek Llama service with a just allowed attack rate of 10 RPM based on the \emph{Transformers} library using 4 NVIDIA 2080ti GPUs with different output token limitations, where only the unrealistic limitation of 128 tokens can successfully defend the attack.}
    \label{fig:attack_6s}
\end{figure*}


\subsection{Attack Results on LLM Service Systems}

We then evaluate attack results of the proposed ThinkTrap on practical LLM service systems.
Specifically, we deploy the DS Llama 8B LLM as a local LLM service on a GPU server equipped with four NVIDIA RTX 2080ti GPUs. To evaluate the system's robustness, we offline-generate 100 adversarial prompts and submit them to the service via its API at a low injection rate of 10 prompts per minute (RPM=10), thereby avoiding detection by standard rate-limiting mechanisms. The maximum number of tokens per generation is set to 32,768 to enable extensive output, which is a common setting of modern LLM service. 
The metrics of time to first token (TTFT), tokens per second (TPS), GPU memory consumption of the GPU server are plotted on Fig. \ref{fig:attack_6s}.

\textbf{ThinkTrap effectively exhausts the computational resources of the LLM server.} The crafted adversarial prompts result in substantially prolonged inference durations. Notably, the system latency increases approximately linearly prior to the 40th input, suggesting that each prompt reliably induces around four minutes of sustained computation. Beyond this point, the rate of latency growth slightly tapers, as some earlier generations complete. Nevertheless, due to the continuous arrival of new prompts and the already degraded computational throughput, the overall latency continues to escalate rapidly.

\textbf{ThinkTrap effectively depletes the GPU memory resources of the LLM server, thereby enabling a successful denial-of-service (DoS) attack.} We further observe a sustained increase in GPU memory consumption across all four devices. Initially, each GPU utilizes no more than 4GB of memory. However, after the injection of 80 adversarial prompts, the most heavily loaded GPU reaches 8GB of usage. This growth is primarily attributed to the accumulation of key-value (KV) caches for each ongoing inference, which demands substantial memory resources. Considering that a portion of GPU memory is reserved for the inference framework itself, this level of consumption approaches the capacity limit of consumer-grade GPUs such as the RTX 2080 Ti. Consequently, after the 80th input, the system experiences memory exhaustion, causing many inference requests to fail or time out, indicating a successful denial-of-service (DoS) attack. While the system remains partially responsive afterward, its processing speed remains severely degraded, and memory usage continues to rise until the next crash occurs.

\textbf{It's hard to defend against ThinkTrap by simply limiting the maximum number of output tokens, as such constraints can significantly degrade service quality.} One naive way to defend against the ThinkTrap attack is to limit the maximum output tokens allowed to output. In this way, the artificially induced high output is limited. However, we can see from Fig. \ref{fig:attack_6s} that, even with a strict output limitation of 256 tokens, the decoding speed metrics, \emph{i.e.}, TTFT and TPS, still slow down significantly. With the output limitation of 128 tokens, the TTFT and TPS can remain stable, which successfully defense the attack of ThinkTrap. However, such a low output length will greatly affect the user experience, which is obviously not feasible. We can also see that the LLM service of a reasonable length limit of 1024 tokens behaves almost the same as the LLM service without length limit. This shows that the proposed ThinkTrap attack can cause great harm to the system even when the service allows a reasonable maximum output length.

\begin{figure*}
    \centering
    \begin{subfigure}{0.32\textwidth}
        \centering
        \includegraphics[height=4.3cm]{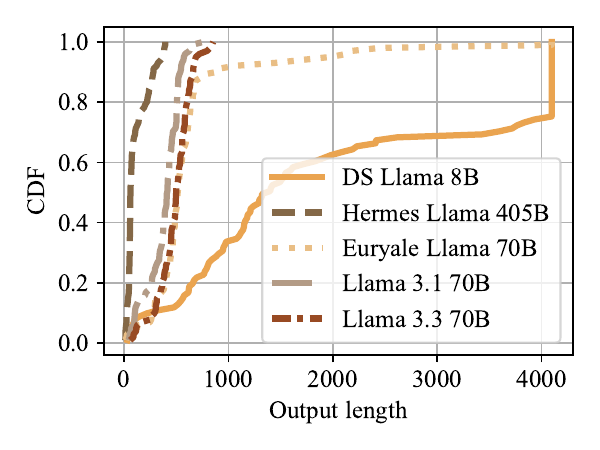}
        \caption{Transfer to the same LLM family}
        \label{fig:cdf_arch}
    \end{subfigure}
    \hfill
    \begin{subfigure}{0.32\textwidth}
        \centering
        \includegraphics[height=4.3cm]{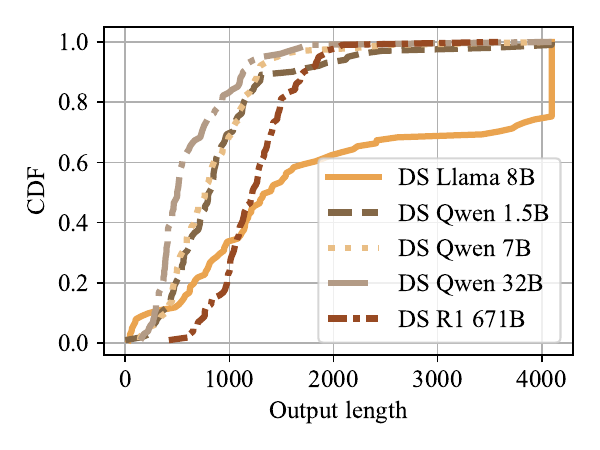}
        \caption{Transfer to the same SFT dataset}
        \label{fig:cdf_data}
    \end{subfigure}
    \hfill
    \begin{subfigure}{0.32\textwidth}
        \centering
        \includegraphics[height=4.3cm]{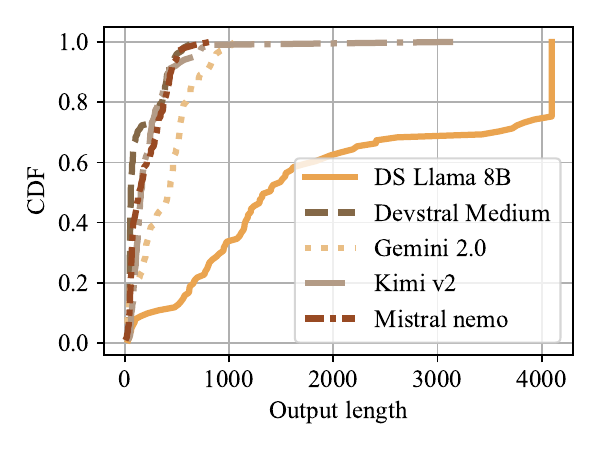}
        \caption{Transfer to brand new LLMs}
        \label{fig:cdf_new}
    \end{subfigure}
    \caption{Cumulative distribution function (CDF) of the output lengths of attack prompts generated for DS Llama 8B and evaluated across various LLMs, indicating that models fine-tuned on the same dataset may share similar vulnerabilities.}
    \label{fig:cdf}
\end{figure*}





\subsection{Transferability of Attack Prompts}

We further evaluate the transferability of the generated attack prompts, given the substantial computational cost associated with their generation, \emph{i.e.}, we assess the feasibility of reusing prompts crafted to exploit one LLM to successfully attack other LLMs without incurring additional generation overhead. To this end, we evaluate the transferability of offline-generated attack prompts for DS Llama 8B as an example.

\subsubsection{Transferring across LLM Families}

We first evaluate the transferability of attack prompt across LLM families. To this end, we plot the output length of attack prompts generated for the DS Llama 8B for several Llama-based LLMs, \emph{i.e.}, Hermes Llama 405B, Euryale Llama 70B, Llama 3.1 70B, Llama 3.3 70B. The cumulative distribution function (CDF) on the output length of the evaluated LLMs, including the source DS Llama 8B, is shown in Fig. \ref{fig:cdf_arch}.

\textbf{The attack prompts exhibit limited transferability even within the same LLM family.} As shown in Fig. \ref{fig:cdf_arch}, even within the same model family, \emph{e.g.}, Llama, the effectiveness of the attack prompts significantly degrades when transferred to another model variant. The resulting output lengths are consistently below 800 tokens, substantially shorter than those observed in the source LLM. This observation suggests that the exploited vulnerabilities are not inherent to the shared architecture or parameter space of the LLM family.

\subsubsection{Transferring across LLM Supervised Fine-Tuning (SFT) Datasets}

We then evaluate the transferability of attack prompts across Supervised Fine-Tuning (SFT) datasets. Specifically, we evaluate the output length of the LLMs fine-tuned on the DeepSeek R1 distilled dataset, \emph{i.e.}, DS R1, DS Qwen 1.5B, DS Qwen 7B, DS Qwen 32B, and DS R1 671B, also on attack prompts generated for DS Llama 8B. The CDF of output lengths is shown in Fig. \ref{fig:cdf_data}.

\textbf{The attack prompts exhibit strong transferability across models fine-tuned on the same supervised fine-tuning (SFT) dataset.} As illustrated in Fig. \ref{fig:cdf_data}, although some degradation in performance is observed, models fine-tuned on the same SFT dataset, \emph{i.e.}, DeepSeek R1, remain largely susceptible to each other’s attack prompts. Notably, prompts generated by the DS Llama model frequently elicit outputs exceeding 4096 tokens, with a high likelihood of surpassing 800 tokens. This suggests that the vulnerability associated with infinite generation is likely introduced during the post-pretraining SFT phase. Furthermore, we observe that the output length of the source DS Llama 8B model does not consistently reach 4096 tokens, which can be attributed to the inherent stochasticity of LLM decoding. Nevertheless, the generated outputs are sufficiently long to preserve the effectiveness of the attack.

\subsubsection{Transferring to Brand New LLMs}

We also evaluate the transferability of attack prompts on the brand new LLMs to access its universal generalization ability. To this end, we evaluate the output lengths of the new LLMs that are unrelated to the DS Llama 8B LLM, \emph{i.e.}, Devstral Medium, Gemini 2.0, Kimi v2, and Mistral nemo. The CDF of output length is shown in Fig. \ref{fig:cdf_new}.

\textbf{The attack prompts demonstrate limited generalizability when applied to unseen LLMs.} As shown in Fig. \ref{fig:cdf_new}, attack prompts crafted specifically for the DS Llama model are largely ineffective against novel LLMs. Despite variations in output lengths across different models, most fail to approach the maximum length of 4096 tokens, with the majority producing outputs of fewer than 800 tokens. This outcome supports the validity of using 4096 as a practical upper bound for evaluating excessive output generation, as non-adaptive prompts rarely trigger such extended outputs. Moreover, the results indicate that the generated prompts lack cross-model generalization, suggesting that each LLM exhibits unique susceptibility patterns and requires tailored attack samples.

\section{Defending Against ThinkTrap}
\label{sec:defense}

In this section, we examine practical defense mechanisms against ThinkTrap and analyze their effectiveness in real-world LLM serving environments. We focus on two representative mitigation strategies that are widely deployed in current LLM hosting systems, namely lightweight anomaly detection and resource-aware scheduling. Our evaluation further investigates their operational implications, providing guidance for practitioners who must balance robustness and service quality.

\subsection{Defense Mechanisms}

We consider the following two typical practical defense mechanisms:

\begin{itemize}
    \item \textbf{Anomaly Detection} \cite{li2023repetition, li2023contrastive}: An anomaly detection mechanism is implemented to identify repetitive or looping generations through an $n$-gram-based analysis of model outputs. Consecutive token sequences, \emph{i.e.}, $n$-grams, are continuously monitored within a sliding window, and requests exhibiting excessive recurrence frequency are regarded as degenerated and terminated early to prevent unnecessary computation. 4-grams are employed \cite{li2023repetition} in our implementation to provide a trade-off between detection sensitivity and robustness against benign stylistic repetition.
    \item \textbf{Resource-aware Scheduling} \cite{sheng2024fairness, zhang2025sgdrc, zhang2024novas}: We implement a resource-aware scheduling mechanism following the Virtual Token Counter (VTC) policy \cite{sheng2024fairness}, which enforces fine-grained control over decoding progress to prevent unbounded resource occupation. Instead of allowing a request to decode continuously, the scheduler allocates each active request a fixed token quantum (\emph{e.g.}, 1024 tokens) per scheduling round. Once this quota is exhausted, the request is preempted, its state is cached, and it is re-queued behind other pending tasks. This token-level preemption ensures that no single request can occupy GPU or NPU resources for an extended interval, thereby limiting the impact of long-generation abuse patterns.
\end{itemize}

\subsection{Effectiveness of Defense Mechanisms}


\begin{figure}
\centering
\includegraphics[width=\linewidth]{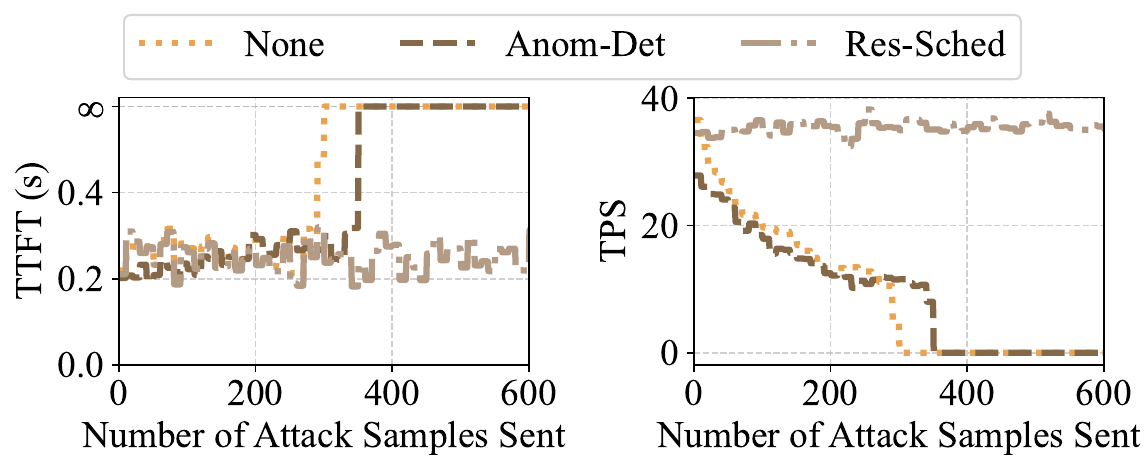}
\caption{Time-to-First-Token (TTFT) and Throughput (TPS) of the LLM service under the ThinkTrap attack (10 RPM) with anomaly detection (Anom-Det) and resource-aware scheduling (Res-Sched). Anomaly detection offers limited protection and adds overhead, while resource-aware scheduling mitigates the attack at the cost of degraded QoS for long-form requests.}
\label{fig:defense}
\end{figure}

We evaulate the defense performance on a server equipped with eight Ascend 910B NPUs, each providing 64 GB of high-bandwidth memory (HBM), is employed for the evaluation. We select DeepSeek Llama 70B as the representative model, given its widespread adoption and the suitability of its scale for this hardware configuration. The model is deployed using MindIE, which schedules decoding requests in a First-In-First-Out (FIFO) manner. The adversarial prompts are pre-generated and subsequently issued to the target server at a controlled rate of 10 requests per minute (RPM).
Fig. \ref{fig:defense} illustrates the Time-to-First-Token (TTFT) and Throughput (TPS) of the LLM service under the ThinkTrap attack with the Anomaly Detection and Resource-aware Scheduling defense mechanisms, respectively.

\textbf{Anomaly detection defense is ineffective against ThinkTrap.} We can see from Fig.~\ref{fig:defense} that although anomaly detection slightly alleviates the denial-of-service (DoS) effect caused by ThinkTrap, its protection capacity remains extremely limited. It can only tolerate several tens of adversarial prompts before the service eventually collapses. This weakness arises because the long-form outputs induced by ThinkTrap are not merely mechanical repetitions of surface tokens but exhibit semantic-level redundancy with moderate diversity, making them difficult to capture through lightweight repetition detectors.
Moreover, the anomaly detection approach incurs a noticeable performance overhead. As shown in Fig. \ref{fig:defense}, its TPS is lower even at the beginning of the attack. This indicates that, despite being computationally lightweight, the detector must still inspect every LLM decoding stream, thereby reducing overall throughput. More sophisticated anomaly detection methods, such as those employing an auxiliary language model to assess semantic repetition, would consume substantially more computational resources, rendering them impractical for real-time inference services.

\textbf{The resource-aware scheduling approach successfully defenses the attack, albeit with a trade-off in service quality for long-response requests.} 
As shown in Fig. \ref{fig:defense}, under single-user attack scenarios, the resource-aware scheduling mechanism effectively mitigates the ThinkTrap attack by constraining the maximum decoding length of each inference request. Once the predefined limit is reached, the decoding process is terminated and the corresponding computational resources are promptly released. This mechanism prevents malicious requests from monopolizing the hardware for unbounded periods and thus protects the service from complete breakdown.
However, this improvement in system stability comes at the cost of degraded quality of service (QoS) for legitimate requests that inherently require long-form reasoning. Tasks requiring long outputs such as mathematical problem solving and embodied task planning experience frequent interruptions and forced re-scheduling, leading to a significant reduction in their overall inference throughput. 
Furthermore, because scheduling operates in a sequential manner, the defense remains vulnerable under concurrent multi-user attacks. 
Such queuing pressure is well known in multi-user video analytics systems \cite{zhang2024blind}. 
When multiple adversarial clients issue long-generation requests simultaneously, the scheduler becomes congested, causing benign requests to experience excessive queuing delays and inflated TTFT.

\subsection{Discussion of Defenses}

Defense evaluation demonstrates that resource-aware scheduling is highly effective in preventing adversarial requests from monopolizing decoding resources, thereby preserving service availability under sustained attack. Such mechanisms, however, inevitably introduce additional latency for benign requests that require long, uninterrupted generations. This tradeoff highlights a fundamental tension in serving large-scale LLMs, where the system must maintain fairness and availability in the presence of adversaries while simultaneously supporting emerging workloads that demand extended reasoning or narrative outputs. We argue that resource-aware scheduling should therefore be treated as a first-class requirement for modern LLM hosting platforms, rather than an optional optimization.

Beyond the mechanisms evaluated in this study, several operational safeguards, such as output-length caps, per-user rate limiting, and per-query compute metering, can limit an attacker’s ability to induce unbounded inference. However, these strategies degrade the experience for legitimate users whose tasks naturally require long outputs or multi-step reasoning. Their role resembles classical DoS mitigation approaches that preserve availability by selectively reducing functionality or imposing flow control under extreme load. While effective as last-resort measures, these approaches are fundamentally coarse-grained and do not address the root cause of reasoning-induced DoS behavior.

Looking forward, more principled defenses will likely require a deeper understanding of the computation pathways exercised during long-form and multi-step LLM reasoning. Developing real-time signals that reflect internal model states, decoding complexity, or incremental resource usage could enable more adaptive scheduling and throttling mechanisms. Attack-aware resource allocation should also be incorporate into LLM-serving frameworks, balancing robustness, throughput, and model utility under adversarial environments.

\section{Conclusion}

This paper investigates a previously overlooked vulnerability in closed-source LLM services: their susceptibility to DoS attacks via adversarial prompts. We propose ThinkTrap, a black-box attack framework that identifies prompts inducing excessive computation by leveraging derivative-free optimization in a continuous surrogate space. Through a low-dimensional, token-wise strategy, ThinkTrap circumvents the challenges of discrete input spaces and high-dimensional optimization. Extensive evaluations on commercial and self-hosted LLMs demonstrate that ThinkTrap can degrade system performance by inflating output length, exhausting GPU resources, and delaying legitimate queries. These results highlight a critical, asymmetric threat to LLM infrastructure and underscore the need for prompt-level defenses in black-box deployment settings. 

\section*{Ethics Statements}

This work investigates prompting-based denial-of-service risks in LLM services. Given the sensitivity of studying system abuse vectors, we followed strict ethical and responsible research practices throughout the research.
All stress experiments were conducted on privately hosted and university-managed LLM deployments, \emph{i.e.}, models deployed on the Zhiyuan-1 cluster maintained by the Center for High Performance Computing at Shanghai Jiao Tong University, with explicit permission from system administrators. 
These environments allowed full instrumentation and stress evaluation to safely examine worst-case execution behavior.

For commercial LLMs (\emph{e.g.}, ChatGPT, Gemini, DeepSeek), we only conducted limited and controlled trial queries during June and July 2025. These queries strictly adhered to publicly documented usage policies, did not exceed normal user behavior patterns, and did not create abnormal load or service disruption. No unauthorized stress tests were performed on commercial infrastructure.
Following reviewer guidance on responsible disclosure, we formally contacted all evaluated LLM providers via their official security reporting channels on October 10, 2025, summarizing the ThinkTrap mechanism, its potential impact, and mitigation insights.
Our goal is to responsibly surface security weaknesses to strengthen AI system robustness, not to enable misuse. We do not release attack-specific configurations that could facilitate abuse, and we encourage LLM platform operators to adopt proactive safeguards. No personal data, user content, or proprietary system logs were accessed in this study.


\section*{Acknowledgment}

This work was supported in part by the Natural Science Foundation of China (Grants No. 62432008, 62472083). 
The computations in this paper were run in part on the \mbox{Zhiyuan-1} cluster supported by the Center for High Performance Computing at Shanghai Jiao Tong University.




%
\balance
\bibliographystyle{IEEEtran}
\bibliography{base.bib}

@article{sanderson2023gpt,
  title={GPT-4 is here: what scientists think},
  author={Sanderson, Katharine},
  journal={Nature},
  volume={615},
  number={7954},
  pages={773},
  year={2023},
  publisher={Nature}
}

@article{mon2025embodied,
  title={Embodied large language models enable robots to complete complex tasks in unpredictable environments},
  author={Mon-Williams, Ruaridh and Li, Gen and Long, Ran and others},
  journal={Nature Machine Intelligence},
  pages={1--10},
  year={2025},
  publisher={Nature Publishing Group UK London}
}

@inproceedings{li2024embodied,
  title={Embodied agent interface: Benchmarking llms for embodied decision making},
  author={Li, Manling and Zhao, Shiyu and others},
  booktitle={Proceedings of NeurIPS},
  year={2024}
}

@article{qiu2024llm,
  title={LLM-based agentic systems in medicine and healthcare},
  author={Qiu, Jianing and Lam, Kyle and Li, Guohao and Acharya, Amish and Wong, Tien Yin and Darzi, Ara and Yuan, Wu and Topol, Eric J},
  journal={Nature Machine Intelligence},
  volume={6},
  number={12},
  pages={1418--1420},
  year={2024},
  publisher={Nature Publishing Group}
}

@article{zheng2025large,
  title={Large language models for scientific discovery in molecular property prediction},
  author={Zheng, Yizhen and Koh, Huan Yee and Ju, Jiaxin and Nguyen, Anh TN and May, Lauren T and Webb, Geoffrey I and Pan, Shirui},
  journal={Nature Machine Intelligence},
  pages={1--11},
  year={2025},
  publisher={Nature Publishing Group UK London}
}

@article{truhn2023large,
  title={Large language models should be used as scientific reasoning engines, not knowledge databases},
  author={Truhn, Daniel and Reis-Filho, Jorge S and Kather, Jakob Nikolas},
  journal={Nature medicine},
  volume={29},
  number={12},
  pages={2983--2984},
  year={2023},
  publisher={Nature Publishing Group US New York}
}

@inproceedings{schuba1997analysis,
  title={Analysis of a denial of service attack on TCP},
  author={Schuba, Christoph L and Krsul, Ivan V and Kuhn, Markus G and Spafford, Eugene H and Sundaram, Aurobindo and Zamboni, Diego},
  booktitle={Proceedings of IEEE S\&P},
  year={1997}
}

@article{pelechrinis2010denial,
  title={Denial of service attacks in wireless networks: The case of jammers},
  author={Pelechrinis, Konstantinos and Iliofotou, Marios and Krishnamurthy, Srikanth V},
  journal={IEEE Communications surveys \& tutorials},
  volume={13},
  number={2},
  pages={245--257},
  year={2010},
  publisher={IEEE}
}

@inproceedings{dong25,
  author       = {Jianshuo Dong and
                  Ziyuan Zhang and
                  Qingjie Zhang and
                  Tianwei Zhang and
                  Hao Wang and
                  Hewu Li and
                  Qi Li and
                  Chao Zhang and
                  Ke Xu and
                  Han Qiu},
  title        = {An Engorgio Prompt Makes Large Language Model Babble on},
  booktitle    = {Proceedings of ICLR},
  year         = {2025}
}

@inproceedings{shumailov2021sponge,
  title={Sponge examples: Energy-latency attacks on neural networks},
  author={Shumailov, Ilia and Zhao, Yiren and Bates, Daniel and Papernot, Nicolas and Mullins, Robert and Anderson, Ross},
  booktitle={Proceedings of IEEE EuroS\&P},
  year={2021}
}

@article{kumar2025overthinking,
  title={OVERTHINKING: Slowdown Attacks on Reasoning LLMs},
  author={Kumar, Abhinav and Roh, Jaechul and Naseh, Ali and Karpinska, Marzena and Iyyer, Mohit and Houmansadr, Amir and Bagdasarian, Eugene},
  journal={arXiv preprint arXiv:2502.02542},
  year={2025}
}

@article{feng2024llmeffichecker,
  title={Llmeffichecker: Understanding and testing efficiency degradation of large language models},
  author={Feng, Xiaoning and Han, Xiaohong and Chen, Simin and Yang, Wei},
  journal={ACM Transactions on Software Engineering and Methodology},
  volume={33},
  number={7},
  pages={1--38},
  year={2024},
  publisher={ACM New York, NY}
}

@article{chen2024not,
  title={Do not think that much for 2+ 3=? on the overthinking of o1-like llms},
  author={Chen, Xingyu and Xu, Jiahao and Liang, Tian and He, Zhiwei and Pang, Jianhui and Yu, Dian and Song, Linfeng and Liu, Qiuzhi and Zhou, Mengfei and Zhang, Zhuosheng and others},
  journal={arXiv preprint arXiv:2412.21187},
  year={2024}
}

@inproceedings{armen21,
  author       = {Armen Aghajanyan and
                  Sonal Gupta and
                  Luke Zettlemoyer},
  title        = {Intrinsic Dimensionality Explains the Effectiveness of Language Model Fine-Tuning},
  booktitle    = {Proceedings of ACL},
  year         = {2021}
}

@article{qin2024exploring,
  title={Exploring Universal Intrinsic Task Subspace for Few-Shot Learning via Prompt Tuning},
  author={Qin, Yujia and Wang, Xiaozhi and Su, Yusheng and Lin, Yankai and Ding, Ning and Yi, Jing and Chen, Weize and Liu, Zhiyuan and Li, Juanzi and Hou, Lei and others},
  journal={IEEE/ACM Transactions on Audio, Speech, and Language Processing},
  year={2024},
  publisher={IEEE}
}

@inproceedings{xu24lora,
  author       = {Yuhui Xu and
                  Lingxi Xie and
                  Xiaotao Gu and
                  Xin Chen and
                  Heng Chang and
                  Hengheng Zhang and
                  Zhengsu Chen and
                  Xiaopeng Zhang and
                  Qi Tian},
  title        = {QA-LoRA: Quantization-Aware Low-Rank Adaptation of Large Language
                  Models},
  booktitle    = {Proceedings of  ICLR},
  year         = {2024}
}

@inproceedings{sun2022black,
  title={Black-box tuning for language-model-as-a-service},
  author={Sun, Tianxiang and Shao, Yunfan and Qian, Hong and Huang, Xuanjing and Qiu, Xipeng},
  booktitle={Proceedings of ICML},
  year={2022}
}

@article{vaswani2017attention,
  title={Attention is all you need},
  author={Vaswani, Ashish and Shazeer, Noam and Parmar, Niki and Uszkoreit, Jakob and Jones, Llion and Gomez, Aidan N and Kaiser, {\L}ukasz and Polosukhin, Illia},
  journal={Proceedings of NeurIPS},
  year={2017}
}

@inproceedings{muller2024impact,
  title={The Impact of Uniform Inputs on Activation Sparsity and Energy-Latency Attacks in Computer Vision},
  author={M{\"u}ller, Andreas and Quiring, Erwin},
  booktitle={Proceedings of IEEE SPW},
  year={2024}
}

@inproceedings{shapira2023phantom,
  title={Phantom sponges: Exploiting non-maximum suppression to attack deep object detectors},
  author={Shapira, Avishag and Zolfi, Alon and Demetrio, Luca and Biggio, Battista and Shabtai, Asaf},
  booktitle={Proceedings of IEEE/CVF WACV},
  year={2023}
}

@inproceedings{schoof2024beyond,
  title={Beyond PhantomSponges: Enhancing Sponge Attack on Object Detection Models},
  author={Schoof, Coen and Koffas, Stefanos and Conti, Mauro and Picek, Stjepan},
  booktitle={Proceedings of ACM WiseML},
  year={2024}
}

@inproceedings{ma2024slowtrack,
  title={Slowtrack: Increasing the latency of camera-based perception in autonomous driving using adversarial examples},
  author={Ma, Chen and Wang, Ningfei and Chen, Qi Alfred and Shen, Chao},
  booktitle={Proceedings of AAAI},
  year={2024}
}

@inproceedings{liu2023slowlidar,
  title={Slowlidar: Increasing the latency of lidar-based detection using adversarial examples},
  author={Liu, Han and Wu, Yuhao and Yu, Zhiyuan and Vorobeychik, Yevgeniy and Zhang, Ning},
  booktitle={Proceedings of IEEE/CVF CVPR},
  year={2023}
}

@inproceedings{chen2022nmtsloth,
  title={Nmtsloth: understanding and testing efficiency degradation of neural machine translation systems},
  author={Chen, Simin and Liu, Cong and Haque, Mirazul and Song, Zihe and Yang, Wei},
  booktitle={Proceedings of ACM FSE},
  year={2022}
}

@inproceedings{rico16,
  author       = {Rico Sennrich and
                  Barry Haddow and
                  Alexandra Birch},
  title        = {Neural Machine Translation of Rare Words with Subword Units},
  booktitle    = {Proceedings of ACL},
  year         = {2016}
}

@inproceedings{geiping2024coercing,
  title={Coercing LLMs to Do and Reveal (Almost) Anything},
  author={Geiping, Jonas and Stein, Alex and Shu, Manli and Saifullah, Khalid and Wen, Yuxin and Goldstein, Tom},
  booktitle={Proceedings of ICLR STLLM Workshop},
  year={2024}
}

@inproceedings{hou2023promptboosting,
  title={Promptboosting: Black-box text classification with ten forward passes},
  author={Hou, Bairu and O’connor, Joe and Andreas, Jacob and Chang, Shiyu and Zhang, Yang},
  booktitle={Proceedings of ICML},
  year={2023}
}

@article{zou2023universal,
  title={Universal and transferable adversarial attacks on aligned language models},
  author={Zou, Andy and Wang, Zifan and Carlini, Nicholas and Nasr, Milad and Kolter, J Zico and Fredrikson, Matt},
  journal={arXiv preprint arXiv:2307.15043},
  year={2023}
}

@inproceedings{XuLDLP24,
  author       = {Zihao Xu and
                  Yi Liu and
                  Gelei Deng and
                  Yuekang Li and
                  Stjepan Picek},
  title        = {A Comprehensive Study of Jailbreak Attack versus Defense for Large Language Models},
  booktitle    = {Findings of ACL},
  year         = {2024}
}

@inproceedings{GuoYZQ024,
  author       = {Xingang Guo and
                  Fangxu Yu and
                  Huan Zhang and
                  Lianhui Qin and
                  Bin Hu},
  title        = {COLD-Attack: Jailbreaking LLMs with Stealthiness and Controllability},
  booktitle    = {Proceedings of ICML},
  year         = {2024}
}

@article{brown2020language,
  title={Language models are few-shot learners},
  author={Brown, Tom and Mann, Benjamin and Ryder, Nick and Subbiah, Melanie and Kaplan, Jared D and Dhariwal, Prafulla and Neelakantan, Arvind and Shyam, Pranav and Sastry, Girish and Askell, Amanda and others},
  journal={Proceedings of NeurIPS},
  year={2020}
}

@article{gurnee2023language,
  title={Language models represent space and time},
  author={Gurnee, Wes and Tegmark, Max},
  journal={arXiv preprint arXiv:2310.02207},
  year={2023}
}

@article{touvron2023llama,
  title={Llama 2: Open foundation and fine-tuned chat models},
  author={Touvron, Hugo and Martin, Louis and Stone, Kevin and Albert, Peter and Almahairi, Amjad and Babaei, Yasmine and Bashlykov, Nikolay and Batra, Soumya and Bhargava, Prajjwal and Bhosale, Shruti and others},
  journal={arXiv preprint arXiv:2307.09288},
  year={2023}
}

@inproceedings{auger2012tutorial,
  title={Tutorial CMA-ES: evolution strategies and covariance matrix adaptation},
  author={Auger, Anne and Hansen, Nikolaus},
  booktitle={Proceedings of ACM GECCO Companion},
  year={2012}
}

@incollection{zhang2016gaussian,
  title={Gaussian distribution},
  author={Zhang, Xinhua},
  booktitle={Encyclopedia of machine learning and data mining},
  pages={1--5},
  year={2016},
  publisher={Springer}
}

@article{wu2024new,
	title={A new era in llm security: Exploring security concerns in real-world llm-based systems},
	author={Wu, Fangzhou and Zhang, Ning and Jha, Somesh and McDaniel, Patrick and Xiao, Chaowei},
	journal={arXiv preprint arXiv:2402.18649},
	year={2024}
}

@misc{mindie2023,
  title        = {MindIE: MindSpore Inference Engine for Ascend AI Processors},
  author       = {{Huawei Technologies Co., Ltd.}},
  howpublished = {\url{https://www.mindspore.cn/mindie}},
  note         = {Accessed: 2025-07-02}
}

@inproceedings{li2023repetition,
  title={Repetition in repetition out: Towards understanding neural text degeneration from the data perspective},
  author={Li, Huayang and Lan, Tian and Fu, Zihao and Cai, Deng and Liu, Lemao and Collier, Nigel and Watanabe, Taro and Su, Yixuan},
  booktitle={Proceedings of NeurIPS},
  year={2023}
}

@inproceedings{li2023contrastive,
  title={Contrastive Decoding: Open-ended Text Generation as Optimization},
  author={Li, Xiang Lisa and Holtzman, Ari and Fried, Daniel and Liang, Percy and Eisner, Jason and Hashimoto, Tatsunori B and Zettlemoyer, Luke and Lewis, Mike},
  booktitle={Proceedings of ACL},
  year={2023}
}

@inproceedings{sheng2024fairness,
  title={Fairness in Serving Large Language Models},
  author={Sheng, Y. and et al.},
  booktitle={Proceedings of USENIX OSDI},
  year={2024}
}

@inproceedings{zhang2025sgdrc,
  title={SGDRC: Software-Defined Dynamic Resource Control for Concurrent DNN Inference on NVIDIA GPUs},
  author={Zhang, Yongkang and Yu, Haoxuan and Han, Chenxia and Wang, Cheng and Lu, Baotong and Li, Yunzhe and Jiang, Zhifeng and Li, Yang and Chu, Xiaowen and Li, Huaicheng},
  booktitle={Proceedings of ACM PPoPP},
  year={2025}
}

@article{zhang2024novas,
  title={Novas: Tackling online dynamic video analytics with service adaptation at mobile edge servers},
  author={Zhang, Liang and Zhu, Hongzi and Fei, Wen and Li, Yunzhe and Zhang, Mingjin and Cao, Jiannong and Guo, Minyi},
  journal={IEEE Transactions on Computers},
  volume={73},
  number={9},
  pages={2220--2232},
  year={2024},
  publisher={IEEE}
}

@inproceedings{zhang2024blind,
  title={The Blind and the Elephant: A Preference-aware Edge Video Analytics Scheduler for Maximizing System Benefit},
  author={Zhang, Liang and Zhu, Hongzi and Li, Yunzhe and Shen, Jiangang and Guo, Minyi},
  booktitle={Proceedings of ICPP},
  year={2024}
}


\end{document}